\documentclass[twocolumn]{aastex62}
\usepackage{apjfonts}
\usepackage{graphicx}
\usepackage{dcolumn}
\usepackage{color}
\usepackage{float}
\usepackage{ulem}

\shorttitle{'Imaka Image Quality}
\shortauthors{Abdurrahman et al.}

\begin{document}

\title{Improved Image Quality Over $10\arcmin$ Fields with the `Imaka Ground Layer Adaptive Optics Experiment }

\author[0000-0002-9915-8195]{F. Abdurrahman}
\affiliation{Department of Astronomy, University of California, Berkeley}

\author[0000-0001-9611-0009]{J. R. Lu}
\affiliation{Department of Astronomy, University of California, Berkeley}

\author{M. Chun}
\affiliation{Institute for Astronomy, University of Hawai`i at M\={a}noa, Hilo, HI 96720-2700, USA} 

\author{M.W. Service}
\affiliation{Institute for Astronomy, University of Hawai`i at M\={a}noa, Hilo, HI 96720-2700, USA}

\author[0000-0001-5656-7346]{O. Lai}
\affiliation{UCA, Observatoire de la C\^{o}te d'Azur, CNRS, Laboratoire Lagrange, 06304 Nice cedex 4, France}

\author[0000-0001-9259-2688]{D. F{\"o}hring}
\affiliation{Institute for Astronomy, University of Hawai`i at M\={a}noa, Honolulu, HI 96822-1839, USA}

\author{D.Toomey}
\affiliation{Mauna Kea Infrared}

\author[0000-0002-1917-9157]{C. Baranec}
\affiliation{Institute for Astronomy, University of Hawai`i at M\={a}noa, Hilo, HI 96720-2700, USA}

\date{\today}

\begin{abstract}
`Imaka is a ground layer adaptive optics (GLAO) demonstrator on the University of Hawaii 2.2m telescope with a $24\arcmin\times18\arcmin$ field-of-view, nearly an order of magnitude larger than previous AO instruments. In 15 nights of observing with natural guide star asterisms $\sim16\arcmin$ in diameter, we measure median AO-off and AO-on empirical full-widths at half-maximum (FWHM) of 0\farcs95 and 0\farcs64 in $R$-band, 0\farcs81 and 0\farcs48 in $I$-band, and 0\farcs76 and 0\farcs44 at 1 micron. This factor of 1.5-1.7 reduction in the size of the point spread function (PSF) results from correcting both the atmosphere and telescope tracking errors. The AO-on PSF is uniform out to field positions $\sim5\arcmin$ off-axis, with a typical standard deviation in the FWHM of 0\farcs018. Images exhibit variation in FWMM by 4.5\% across the field, which has been applied as a correction to the aforementioned quantities. The AO-on PSF is also $10\times$ more stable in time compared to the AO-off PSF. In comparing the delivered image quality to proxy measurements, we find that in both AO-off and AO-on data, delivered image quality is correlated with `imaka's telemetry, with $R$-band correlation coefficients of 0.68 and 0.70, respectively. At the same wavelength, the data are correlated to DIMM and MASS seeing with coefficients of 0.45 and 0.55. Our results are an essential first step to implementing facility-class, wide-field GLAO on Maunakea telescopes, enabling new opportunities to study extended astronomical sources, such as deep galaxy fields, nearby galaxies or star clusters, at high angular resolution. 
%
%
\end{abstract}

\section{\label{sec:intro}Introduction}
Adaptive optics (AO) is a powerful tool for correcting distortions caused by Earth's atmosphere and delivering diffraction-limited images. Despite the advantages they present, most AO systems have a relatively small field-of-view (FoV), which can limit the science applications. For first generation, ``classical'', single-conjugate adaptive optics (SCAO), that use a single guide star and a single deformable mirror (DM) to correct for turbulence at all heights in a single direction, the correction is limited to regions close to the guide star, resulting in diffraction-limited images spanning only a few arcseconds at optical wavelengths and tens of arcseconds in the near-infrared. Because of this, extended sources such as nearby galaxies or nearby star clusters cannot easily be observed with SCAO in their entirety.  

The second generation of AO seeks to mitigate the limitations of SCAO systems in a variety of ways. Multi-conjugate adaptive optics (MCAO) systems use multiple DMs to correct atmospheric turbulence in more than one dimension. Correction on fields of $1\arcmin-2\arcmin$ with angular resolutions of $80-150$ milliarcseconds (mas) have been achieved by MCAO instruments such as GeMS on Gemini \citep{doi:10.1093/mnras/stt2054} and MAD on the VLT \citep{marchetti2007sky}. Multi-object adaptive optics (MOAO) systems that use tomographic reconstruction of several widely separated guide stars ($3\arcmin - 5\arcmin$) to correct small sub-fields of a few arcseconds in size have achieved image resolutions of $\sim$150 mas. On-sky demonstrations of MOAO include Raven \citep{andersen2012status}, Canary \citep{gendron2011moao} and the future MOSAIC instrument, which includes a multi-object spectrograph with MOAO correction \citep{sharples2000mosaic}.

Ground layer adaptive optics (GLAO), first proposed by \cite{Rigaut2002} to improve wide-field imaging for large telescopes, attempts to stretch the limits of AO even further. GLAO uses several guide stars to correct the shared ground layer turbulence of a large field of view, rather than along a single path through the atmosphere, at the expense of not correcting turbulence at higher altitudes. Previous GLAO demonstrations have shown successful image correction for fields of $1\arcmin - 4\arcmin$ with resolutions of $\sim$300 mas, a factor of 2 improvement over the seeing \citep{baranec2009, Hart2010, LBT_glao_2016, 2017Msngr.168....8A}.
SAM on the SOAR telescope is a facility GLAO instrument, achieving free-atmosphere limited images in $I$-band with a resolution of $\sim$500 mas over a $3\arcmin$ FoV \citep{Tokovinin2016}. Similarly, ARGOS on the LBT is expected to improve spatial resolution by a factor of two across a $2\arcmin$ FoV \citep{gassler2012status}. ESO's Adaptive Optics Facility has implemented multiple systems with GLAO correction on VLT telescopes, including GRAAL ($7\arcmin$x$7\arcmin$ in $J$, $H$, and $K$ bands) and the wide field mode on GALACSI-MUSE ($1\arcmin$x$1\arcmin$ in the optical regime) \citep{kuntschner2012operational}. The potential for even larger fields exists for instruments such as ULTIMATE-SUBARU, which expects uniform point spread functions (PSFs) with GLAO correction across a field of $20\arcmin$ \citep{hayano2014ultimate}. 

`Imaka is a GLAO demonstrator on the University of Hawaii 2.2-meter telescope on Maunakea aimed at quantifying the potential of the site for GLAO. The full `imaka 24\arcmin$\times$18\arcmin \ field is an order of magnitude larger than its MCAO predecessors and is the largest GLAO field of view (FoV) deployed to date. Simulations for `imaka predicted a PSF full width at half maximum (FWHM) of $\sim$300 mas \citep{chun2016imaka} in the near-infrared, which is larger than the FWHM delivered by SCAO systems that can reach $\sim$30 mas at similar wavelengths \citep{wizinowich2000first}. As such, GLAO should be thought of not as a replacement for high-fidelity wavefront correction, small FoV AO systems, but as a distinct, wide-field, ``super-seeing'' system with a different set of advantages and science applications. 

The science made possible with GLAO improvements are numerous. The wide field of view of `imaka allows large, crowded stellar fields to be imaged at higher resolution than normally allowed by the natural seeing. The increased acuity over wider fields can improve astrometric measurements significantly, providing more candidates for reference stars. The observation of large resolved objects such as nearby galaxies, with stable, highly corrected PSFs will also allow more detailed study of topics such as galaxy morphology.  Besides creating sharper images, the increased encircled energies obtained by GLAO correction produce a higher signal-to-noise ratio (SNR) for an otherwise identical exposure, a significant advantage for imaging as well as spectroscopy.  Furthermore, `imaka demonstrates the promising possibility of GLAO on a wide variety of telescopes of different sizes on Maunakea, including ten-meter class telescopes such as Keck as well as the forthcoming generation of extremely large telescopes.

Commissioning of `imaka began in October 2016. Analysis of the data from the three subsequent GLAO obsering runs in 2017 are presented in this paper. We first present a brief description of the instrument (\S2). Then we describe the different data sets used and their observations (\S3), and their reduction (\S4). We then define metrics used to characterize these data, including PSF modeling (\S5). Analyses of the GLAO performance, including an examination of wavelength dependence and field variability, are presented (\S6) before we finally discuss the implications of our findings (\S7).

\section{\label{sec:inst}Instrument Description}

\begin{figure}[h]
    \centering
    \includegraphics[width=0.483\textwidth]{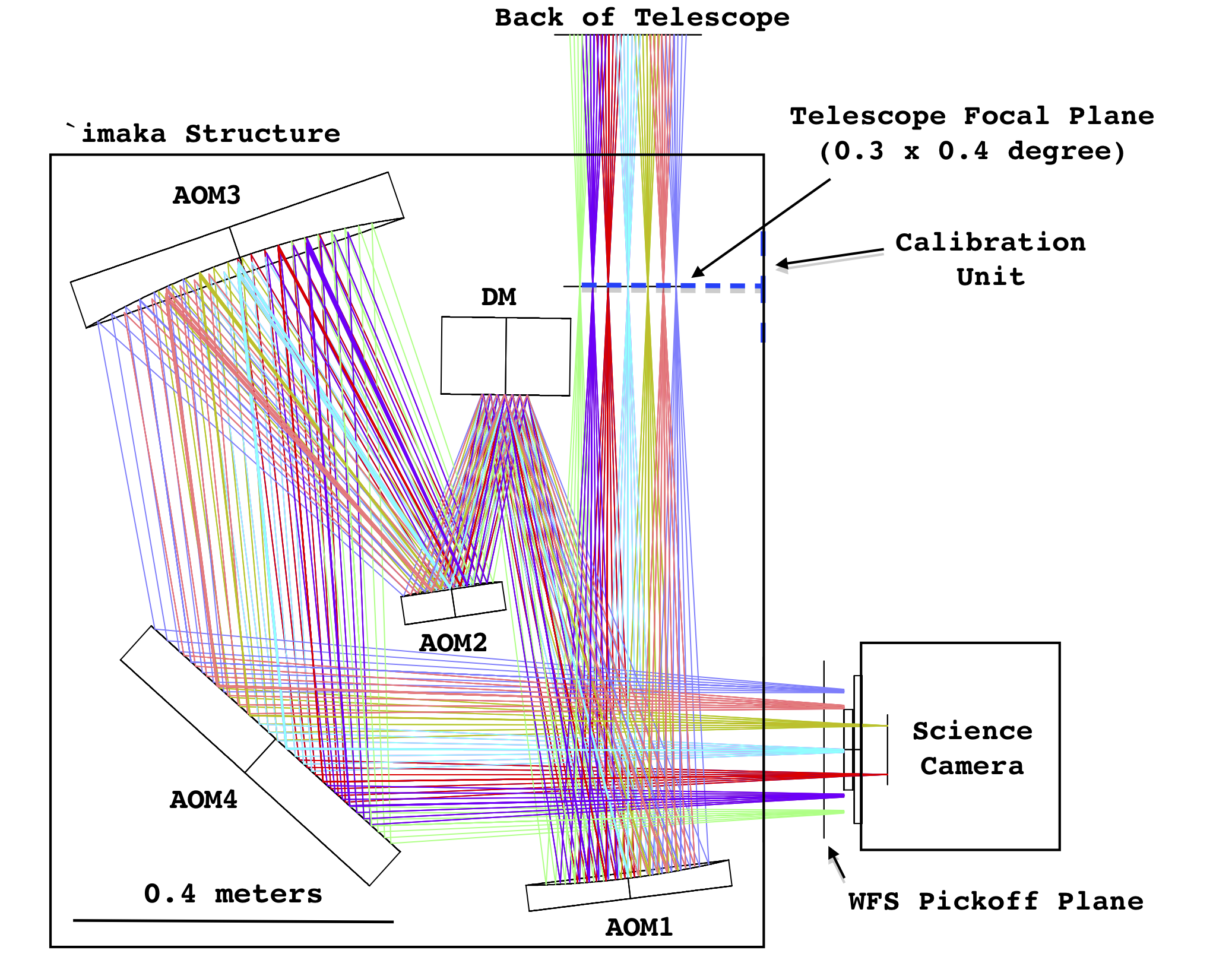}
    \caption{Layout of main elements of the `imaka instrument.  The light from the telescope enters from the top of the diagram.  The first powered element (AOM1) creates a pupil image where the deformable mirror (DM) is placed.   Following the DM are the Offner secondary (AOM2) and the final concave sphere (AOM3).  A large flat (AOM4) folds the beam to a side exit port where the wavefront sensors and science camera are mounted.}
    \label{fig:Layout}
\end{figure}

`Imaka is a demonstrator of wide-field GLAO with its primary goal of providing an on-sky facility to gain experience using and optimizing the performance of GLAO systems. `Imaka is mounted at the Cassegrain focus of the University of Hawaii 2.2 meter telescope on Maunakea.   The instrument is built around a modified Offner optical relay (e.g., \cite{2007SPIE.6691E..06B}) with input and output focal ratios of f/10 and f/13.25 respectively. A schematic layout of the system is shown in Figure $\ref{fig:Layout}$. This optical relay provides a flat pupil position just before the second mirror (AOM2) where we place a curvature deformable mirror consisting of 36 elements.  Note that the deformable mirror is conjugate to the telescope primary mirror (i.e., the ground).  All powered optical elements in the system are spheres, which simplifies alignment and fabrication.  An ``exit'' port near the relayed focal plane provides mounting points for the wavefront sensors and the science camera.

We place the wavefront sensors and the science camera at the exit port of the instrument where an oval field 24\arcmin $\times$18\arcmin ~is passed through the system. We can place up to five Shack-Hartmann wavefront sensors anywhere within this field but we generally observe fields with guide stars within the periphery of the field to avoid vignetting the science camera.    Each wavefront sensor consists of its own optics and camera (Raptor Photonics Kite EMCCD cameras) configured to provide 8$\times$8 sub-apertures with 0\farcs4 pixel$^{-1}$ sampling and a sampling rate up to 180 Hz.  To date, the guide stars used are brighter than R $\leq$ 8.5 mag and we achieve a 0-dB rejection bandwidth of about 15 Hz.  

The science focal plane is located just behind the wavefront sensors. Image quality from the optical relay  was optimized within an approximately 11\arcmin$\times$11\arcmin ~central field and the instrument can accommodate a large 11\arcmin$\times$11\arcmin ~CCD camera or a 7\arcmin$\times$7\arcmin ~near-infrared H4RG-15 camera.  Both of these cameras are still in development so all images to date were taken with a Finger Lakes Instrumentation ML50100 CCD camera.  The camera is mounted slightly off the center of the `imaka science field and can be mounted in a variety of positions in order to mosaic a larger field of view.  The FLI detector consists of 8176$\times$6132, 6$\mu$m pixels with a platescale of 40 mas pixel$^{-1}$ at the `imaka focal plane. The field of view of an single image from the camera was limited by a band-pass filter mounted to the front of the camera to 5\farcm4 $\times$ 4\farcm1.  

A summary of the instrument specifications is provided in Table \ref{table:imakaspecs} and further details can be found in \citep{chun2016imaka} and the references therein.

\begin{deluxetable}{lp{1.9in}}
	\tabletypesize{\footnotesize}
	\tablewidth{0pt}
	\tablecaption{`imaka AO specifications}
	\startdata
    & \\[0.1in]
Telescope & University of Hawaii 2.2m, Maunakea \\
\hspace{0.05in} \\
\hline
\sidehead{Deformable Mirror}
\hspace{0.1in} Type & 	36-element, CILAS curvature mirror\\
\hspace{0.1in} Pupil Size & 57.5 mm pupil \\
\hspace{0.05in} \\
\hline
\sidehead{Wavefront Sensors}
\hspace{0.1in} Number & up to five Shack-Hartmann WFSs \\
\hspace{0.1in} Patrol field & 24\arcmin$\times$18\arcmin\\
\hspace{0.1in} Pupil Sampling & 8x8 subaperture \\
\hspace{0.1in} Pixel size & 0.4\arcsec/pixel \\
\hspace{0.1in} Subaperture Field  &  4.8\arcsec/subaperture \\
\hspace{0.05in} \\
\hline
\sidehead{Wavefront Sensor Cameras}
\hspace{0.1in} Type & Raptor Photonics Kite EMCCD \\
\hspace{0.1in} Wavelengths & $\lambda$ = 0.4-0.7 $\mu$m \\
\hspace{0.1in} Quantum Efficiency & peak 50\%\\
\hspace{0.1in} Readout time & 4.2msec\\
\hspace{0.05in} \\
\hline
\sidehead{System}
\hspace{0.1in} Sampling Rate & up to 180Hz\\
\hspace{0.1in} Rejection Bandwidth & $f_{0dB}\sim 10-15$ Hz\\
 \label{table:imakaspecs}
	\enddata
\end{deluxetable}

\section{\label{sec:obs}Observations}
The data represented in this paper spans three `imaka observing runs: 2017 January 10-14 UT, 2017 February 13-17 UT, and 2017 May 17-22 UT. Data from 2017-01-13 UT were omitted in the analysis due to the MASS instrument not running that night. Our analysis combines three different types of observations, described below\footnote{All of the data from `imaka commissioning are available upon request.}

\subsection{\label{sec:obs_focal_plane}`Imaka Focal Plane Images}
\begin{figure}[h]
    \centering
    \includegraphics[width=0.483\textwidth]{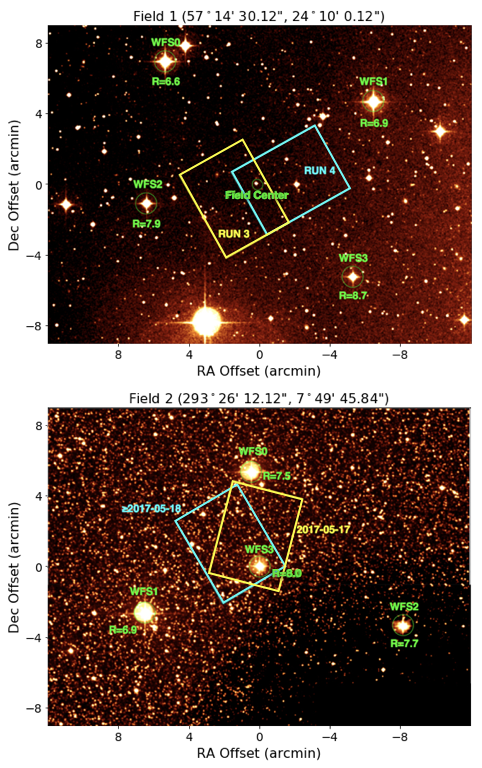}
    \caption{Field 1 (top; 2017 January and 2017 February) and Field 2 (bottom; 2017 May).  In each field, all guide stars used are marked with wave front sensor number and $R$-band brightness.  In Field 1, the east Pleiades field is shown in the yellow rectangle labeled RUN 3, while the west Pleiades field is shown in blue and labeled RUN 4. Positions 1 and 2 of Field 2 are shown in yellow and blue, respectively. In both figures, North is up and East is to the left.  Background images retrieved from 2MASS \citep{skrutskie2006two}.}
    \label{fig:Fields}
\end{figure}

The fields' positions and guide star coordinates for each run are given in Table \ref{tab:coords}.  The two fields used are shown in Figure $\ref{fig:Fields}$.  For the 2017 January and 2017 February runs, we observed the Pleiades star cluster (Field 1). This field has four guide stars, each positioned around the periphery of the science field. For the 2017 May run, we switched to a second field in the galactic plane (Field 2) in order to observe a higher density of stars. This field differed slightly in that it had its fourth guide star in the center of the field, though this star was not always used in the correction (refer to Table \ref{tab:log}). For Field 2, three different guide star and WFS configurations were tested: all four (4WFS) and three outer WFS 0, 1 and 2 (3WFS). For further technical details, refer to \citep{chun2016imaka}.

In order to quantify the gains of `imaka, we made observations both with GLAO correction ({\em AO-on}) and with no correction ({\em AO-off}). To capture both modes in a range of seeing conditions, we cycled through different modes while taking data, (i.e., one image with no correction, one image with GLAO correction, then back to no correction). Rather than either freezing or resetting when the mode switched from AO-on to AO-off, the DM was set to an average of the last 5-10 seconds at the beginning of each AO-off exposure. On average, each night produced 36 frames in each mode. 

Science images were taken with exposure times of 30-60 seconds; these times were chosen in order to average over the seeing. We used $I$-band and $R$-band Johnson filters (centered at 806 nm and 658 nm, respectively), and a ``long-pass'' filter with an effective wavelength of approximately 1 micron. Image position and filter information is listed for each night in Table \ref{tab:log}. In addition to our science images, we took dithered sky images throughout each night, as well as twilights at the beginning of each run for use as flat fields.

\begin{deluxetable}{lcccc}[h!]
	\tabletypesize{\footnotesize}
	\tablewidth{20pt}
	\tablecaption{Field and Guide Star Positions \label{tab:coords}}
	\tablehead{
		Label &
		Name &
		RA (J2000) &
		Dec (J2000) &
		R mag
	}
	\startdata
    \hline
    \multicolumn{5}{c}{Field 1 (January, February)}\\
    \hline
	Field Center & - & 3h48m58.0s & 24$^{\circ}$10'00.1\arcsec & - \\ 
	WFS0 & HD 23873 & 3h49m21.7s & 24$^{\circ}$22'54.4\arcsec & 6.6 \\ 
	WFS1 & HD 23763 & 3h48m29.6s & 24$^{\circ}$20'48.9\arcsec & 6.9 \\
	WFS2 & HD 23886 & 3h49m25.9s & 24$^{\circ}$14'51.7\arcsec & 7.9 \\
	WFS3 & HD 2377 & 3h48m34.8s & 24$^{\circ}$10'52.3\arcsec & 8.7 \\
    \hline
    \multicolumn{5}{c}{Field 2 (May)}\\
    \hline
	Field Center & - & 19h33m44.8s & 7$^{\circ}$49'45.8\arcsec & - \\ 
	WFS0 & HD 184362 & 19h33m46.5s & 7$^{\circ}$55'05.0\arcsec & 7.5 \\ 
	WFS1 & HD 184451 & 19h34m11.1s & 7$^{\circ}$47'08.8\arcsec & 6.9 \\
	WFS2 & HD 184244 & 19h33m12.1s & 7$^{\circ}$46'22.0\arcsec & 7.7 \\
	WFS3 & HD 184336 & 19h33m44.8s & 7$^{\circ}$49'45.7\arcsec & 8.0 \\
	\enddata
    \label{table:coords}
\end{deluxetable}

\begin{deluxetable*}{lccccccccccc}[!]
	\tabletypesize{\footnotesize}
	\tablewidth{0.9\textwidth}
	\tablecaption{Observing Runs\label{tab:log}} 
	\tablehead{
		Date & 
        Field & 
        PA & 
        Filter & 
        Frames &
        AO-Off & 
        AO-On & 
        DIMM$_\lambda$ &
        MASS$_\lambda$ & 
        DIMM$_{500}$ &
        MASS$_{500}$ & 
        Comment \\
        (UT) &  & ($^\circ$) &  & & (mas) & (mas) & (mas) & (mas) & (mas) & (mas) &  
	}
	\startdata
	2017-01-10 & 1 & 119 & R & 20 &        $535\pm17$ & $313\pm20$ & $426\pm7$   & $164\pm2$  & $450\pm8$  & $173\pm2$ & \\
	2017-01-11 & 1 & 119 & R & 212 &       $931\pm20$ & $645\pm10$ & $658\pm13$  & $643\pm18$ & $695\pm14$ & $680\pm19$ & \\
	2017-01-12 & 1 & 119 & R & 157 &       $906\pm13$ & $628\pm7$  & $910\pm16$  & $701\pm18$ & $962\pm17$ & $741\pm19$ &\\
    2017-01-14 (a) & 1 & 119 & R & 61 &    $621\pm15$ & $444\pm14$ & $681\pm17$  & $560\pm28$ & $720\pm18$ & $592\pm30$ & \\
    2017-01-14 (b) & 1 & 29 & I & 84 &    $790\pm13$ & $587\pm11$ & $710\pm16$  & $414\pm12$ & $720\pm18$ & $455\pm14$ & \\
    2017-02-14 & 1 & 29 & I & 56 &         $589\pm17$ & $433\pm12$ & $522\pm12$  & $220\pm22$ & $574\pm13$ & $242\pm24$ & \\
    2017-02-15 & 1 & 29 & I & 45 &         $634\pm42$ & $411\pm24$ & $561\pm24$  & $184\pm5$  & $618\pm27$ & $202\pm6$ & \\
    2017-02-16 & 1 & 29 & I & 14 &         $896\pm28$ & $498\pm17$ & $1006\pm91$ & $160\pm8$  & $1106\pm100$ & $176\pm9$ & \\
    2017-02-17 & 1 & 29 & 1 $\mu m$ & 59 & $663\pm52$ & $377\pm23$ & $445\pm11$  & $297\pm21$ & $511\pm12$ & $341\pm24$ & \\
    2017-02-18 (a) & 1 & 29 & R & 13 &     $552\pm22$ & $370\pm21$ & $220\pm25$  & $112\pm6$  & $233\pm27$ & $118\pm6$ & \\
    2017-02-18 (b) & 1 & 29 & I & 10 &     $661\pm66$ & $321\pm8$  & $193\pm30$  & $110\pm6$ & $213\pm33$ & $121\pm6$ & \\
    2017-05-17 & 2 & 76 & I & 28 &         $882\pm14$ & $332\pm4$  & $368\pm12$  & $112\pm3$  & $405\pm13$ & $124\pm3$ & 4WFS \\
    2017-05-18 & 2 & 120 & I & 68 &        $587\pm9$  & $398\pm2$  & $393\pm9$   & $208\pm5$  & $433\pm9$  & $229\pm5$ & 3WFS \\
    2017-05-19 & 2 & 120 & I & 109 &       $1078\pm25$ &$535\pm2$  & $581\pm11$  & $399\pm11$ & $639\pm12$ & $439\pm12$ & 3WFS \\
    2017-05-20 & 2 & 120 & I & 55 &        $712\pm9$ & $4290\pm2$  & $676\pm16$  & $261\pm6$  & $744\pm18$ & $288\pm7$ & 3WFS \\ 
    2017-05-21 & 2 & 120 & I & 109 &       $644\pm5$ & $368\pm1$   & $540\pm12$  & $115\pm3$  & $595\pm13$ & $127\pm4$ & 3WFS \\
	\enddata
    \tablecomments{Observation setup, conditions, and results summarized with nightly averages. Fields and position angle (PA; measured East of North to the positive y-axis of the detector) refer to those outlined in Figure \textbf{$\ref{fig:Fields}$} and Table \ref{tab:coords}. The reported FWHM is the minor FWHM of a Moffat profile.  As MASS/DIMM seeing is measured at 500nm, we present both a converted DIMM$_\lambda$ and MASS$_\lambda$ that are scaled with equation \ref{eqn:wave_cal} to match the observation wavelengths, in addition to the raw data in the DIMM$_{500}$ and MASS$_{500}$ columns. The nights of 2017-01-14 and 2017-02-18 UT each have two entries, (a) and (b), for the two half-nights in which we observed with two different filters. In the May run, the combination of wavefront sensors changed from all four to only the outer three with different configurations; this change is indicated in the rightmost column.}
	\label{table:log}
\end{deluxetable*}

\subsection{\label{sec:obs_telemetry}`Imaka Telemetry}
In order to evaluate 'imaka performance, we also recorded the AO system wavefront sensor measurements and deformable mirror commands. This data set (the ``telemetry'') provides a means to independently estimate the uncorrected aberrations, the level of correction, and the atmospheric seeing. The telemetry data is taken concurrently with our science images; however, the total recording time for the telemetry stream is limited to $\sim$10 seconds, while the science exposures are 3-6 times longer. The WFS telemetry data is taken without a filter and has an effective wavelength similar to a V+R-band filter. 

The telemetry is used to reconstruct the seeing at a range of altitudes using SloDAR techniques \citep{wilson2002slodar}, although turbulence profiles are not used in this paper. Instead, the telemetry is used to simply derive the integrated, ground layer and free-atmosphere seeing independently from MASS/DIMM. Telemetry data were acquired with no AO correction and are used to compute the slope covariance matrices and maps. These are global in the sense that the covariances are computed across different wavefront sensors (the cross-covariances) as well as with themselves (the auto-covariances). A full description of the telemetry data and its analysis is described in \citep{Lai2018}. The result of the telemetry processing is the estimated seeing for the ground-layer and the free-atmosphere for both AO-off and AO-on images.  



\subsection{\label{sec:obs_mass_dimm}MASS/DIMM}

In addition to our own measurements, we use the seeing reported by the Mauna Kea Weather Center\footnote{\url{http://mkwc.ifa.hawaii.edu}} (MKWC). The MKWC reports seeing measured by the Mauna Kea seeing monitor (MKAM) installed on a seven-meter tall tower between the Canada-France-Hawaii telescope and the Gemini-North telescope. MKAM is approximately 150 meters North of the UH88" telescope.  The seeing monitor consists of a Multi-Aperture Scintillation Sensor (MASS) and a Differential Image Motion Monitor (DIMM) \citep{tokovinin2002differential}. While the DIMM measures integrated seeing of the whole atmosphere, the MASS generates a profile of seeing at a range of altitudes (0.5, 1, 2, 4, 8, and 16 km).  Note that the MKWC MASS data is processed with version v2.047 of the MASS software.  However, there are improvements to the MASS data analysis \citep{lombardi2015using} and we have reprocessed the data using the most recent version of the software \citep{atmos3page}. Note that the data reprocessing still uses the MKAM MASS calibrations from 2009. Throughout this work, we will refer to the MASS seeing as the integrated seeing in the entire MASS profile (e.g. from 500 m to 16 km). 

Both instruments report seeing at 500 nm throughout the night. Every night of focal plane data has corresponding MASS/DIMM measurements with the exception of 2017-01-13 UT when the MASS was not working. We exclude this night from the rest of our analysis. 

It should be noted that the MASS/DIMM instruments do not necessarily observe the same atmospheric profile as `imaka. Ideally, the telemetry's integrated seeing should be comparable to the DIMM, while the free atmosphere seeing should match the MASS. However, the MASS/DIMM points close to zenith, while we observe at a range of altitudes. The differing paths through the atmosphere may result in different turbulence profiles. Additionally, the instruments are located at different places on the summit of MaunaKea and therefore experience different local seeing (e.g., seeing very close to the ground at the seeing monitor and dome seeing at the UH 2.2 m telescope).  

\section{\label{sec:image_reduce}Image Reduction}
Standard image processing was applied to the science images including flat-fielding, sky subtraction, and bad-pixel and cosmic ray rejection. The flat-field image was created from several twilight exposures with different integration times (2-30 seconds) that were normalized by dividing by the $\sigma$-clipped median of all pixels in the image and then combined using a $\sigma$-clipped median. In some cases where twilight images were not taken on a given night, the flat generated from the preceding night was used. Sky images were taken periodically through the observing run each night, usually with a total of about 20 frames. Like the flats, all sky images of a given night were combined using a $\sigma$-clipping median routine that, along with a dithering procedure, removed any stars from the final combined sky image.

In addition to individual science frames, we created stacked images.  For each night, one stacked image was created using the reduced images for each position, control matrix, and filter (e.g., all AO-on images on 2017-05-17 UT taken at \textit{I}-band using the combination of wavefront sensors 0, 1 and 3 were combined to form one image). The frames were matched and had relative astrometric transformations applied before adding them together to create single images, without further correction (e.g., for distortion, differential refraction, etc). Because these stacked images do not capture the effects of seeing variations or zenith distance which are changing between frames, they are used only to examine the variability in AO correction across the field due to the instrument. As such, data from stacked images are only presented in \S6.4 on field variability.

\section{\label{sec:image_quality}Image Quality Analysis}
After reducing the science images, each frame was run through our star finding algorithm. Initial FWHM estimates were obtained by eye from the cleaned science images, typically 0.4\arcsec\space in AO-on and 1.0\arcsec\space in AO-off images.  These initial guesses were used in DAOphot's DAOstarfinder \citep{Stetson1987}, to locate stars in the images.  The star finder performed two passes with this initial PSF guess, locating sources above four standard deviations of the mean pixel value of the image and updating the FWHM guess accordingly.  The images of the Pleiades field yielded about 15 sources each (so few because of the sparseness of the field, the short exposure times, the rejection of saturated sources, and the high SNR needed for the subsequent fitting), while the second, much denser field of the May run had 300-500 sources each. 
%
%

Once the sources were located, certain cuts were made to the sample of stars. First, stars within 20 pixels of image edges or other stars were removed. Second, saturated stars (defined as anything with a peak above 20,000 ADU) were removed. Finally, we defined a minimum flux threshold for each image. Generally, this cut saved only the brightest 80\% of stars in a frame. However, for Field 1, which had few stars to begin with, no flux cut was made.

Once a clean star sample was defined, we characterized the PSF of each star in a given frame using several metrics described below. To further reduce our data, all of our statistics described below were median combined for all stars in a given image, resulting in a single measurement of a given metric for each frame. Thus, each data point in all subsequent plots represents a single frame unless otherwise stated. Note that we did not apply a correction for zenith distance of observations; Figure $\ref{fig:airmass}$ shows our FWHM data as a function of airmass fit to both a constant FWHM and a 3/5 power law, and illustrates that no or very little airmass dependence is seen in our data.

\subsection{\label{sec:psf_models_non_param}Non-Parametric PSF Models}
\begin{figure}[h]
    \centering
    \includegraphics[width=0.45\textwidth]{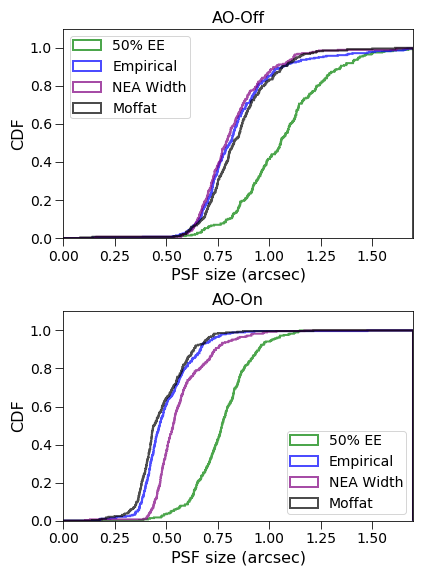}
    \caption{Cumulative distribution functions illustrating different PSF metrics, including the 50\% encircled energy diameter, empirical FWHM, NEA width, and Moffat FWHM.  Though they all approach the PSF characterization slightly differently and thus yield different results, we expect them to be somewhat consistent in the range of values measured on a given dataset. \textit{Top:} AO-off. \textit{Bottom: }AO-on. Note that only $I$-band data is shown here; the four metrics in other filters demonstrate similar differences.}
    \label{fig:Metrics}
\end{figure}

Before establishing a parametric model for the image plane PSFs, we took a numerical approach to their characterization using three different metrics, described below. Each metric quantifies the PSF shape and size  slightly differently; thus the values vary for the same image and can be seen in Figure $\ref{fig:Metrics}$.

{\em Empirical Full Width Half Max: }
Pixels were counted in the vicinity of a star's centroid with ADU counts above half of the maximum pixel value as a brute-force method of calculating a star's FWHM. Assuming this area is circular, we calculated the corresponding diameter: the empirical full width half max.

{\em Encircled Energy Diameter:}
To obtain a more detailed profile of the sources' PSFs, we also calculated encircled energy (EE) diameters for our sources. To do this, we created brightness profiles of each source, counting flux contained within a series of concentric annuli centered on the source's centroid. From this profile, we measured the diameter within which 25\%, 50\%, and 80\% of the total flux was enclosed. 

{\em Noise Equivalent Area: }
An additional way we measure the PSF is a ``noise equivalent area'' (NEA).  A full derivation of the NEA of an arbitrary PSF can be found in \citep{king1983accuracy}, but simply put, it is the area of a region centered on a star's centroid in which the signal-to-noise ratio (SNR) is unity. This is calculated with a similar method to the encircled energy radius, in which concentric circular regions of increasing size around a star are isolated until reaching an SNR of 1. The NEA of a Gaussian profile, for example, is $4\pi \sigma ^2$. For the sake of comparison to PSF measurements that are widths rather than areas, we define the 'NEA width' to be the diameter corresponding to a circle with area NEA.

\subsection{\label{sec:psf_models}Parametric PSF Modeling}
Though the above metrics provide a good sense of image quality, they are not a complete description of stars' PSFs.  These models ignore the shape of the source, which is problematic, as our sources are not radially symmetric. This is particularly apparent in AO-off images, where stars show significant elongation. As such, we parametrically modeled the sources' shape.

We tried a variety of models, including bivariate Gaussian, Lorentz, Moffat, and sinc distributions. Additionally, we tried two component models, including combinations (e.g. Gaussian + Moffat). For each model, the goodness of fit was quantified with the fraction of variance unexplained: 
%
%
\begin{equation}\label{FVU}FVU=\frac{\sum ^{N_{pix}}_i [PSF_{obs,i}-PSF_{mod,i}]^2}{\sum ^{N_{pix}}_i [PSF_{obs,i} - \overline{PSF_{obs}}]^2}\end{equation}

This value was calculated in a box centered on the star with a side length of about 4 times the FWHM. Using this quantity to compare different models, we found that for AO-off images, the best model was a single component elliptical Moffat profile.  We had predicted that for the AO-on images, a two component elliptical Moffat would be ideal; a narrower, brighter component representing the corrected portion of the final image, and a dimmer component with a size similar to that of the AO-off image, representing the uncorrected portion of the PSF.  However, as even a carefully constrained double component model showed no improvement in the FVU, we ultimately chose a single component Moffat for AO-on images as well. 

The model for a two dimensional elliptical Moffat with rotation is defined as:
\begin{equation}
\label{mof1}f(x,y)=a[1+A(x-x_o)^2+B(y-y_o)^2+C(x-x_o)(y-y_o)]^{-\beta}
\end{equation}
\begin{equation}\label{mof2}
A=\left(\frac{\cos\phi}{\alpha_x}\right)^2+
\left(\frac{{\sin}\phi}{\alpha_y}\right)^2, \;
B=\left(\frac{\sin\phi}{\alpha_x}\right)^2+
\left(\frac{{\cos}\phi}{\alpha_y}\right)^2
\end{equation}
\begin{equation}\label{mof3}C=2\sin\phi\cos\phi\left(\frac{1}{\alpha_x}-\frac{1}{\alpha_y}\right)\end{equation}

\noindent
where the star's centroid is given by $x_o$ and $y_o$, the amplitude is $a$, $\beta$ describes the slope of the PSF, $\alpha_x$ and $\alpha_y$ describe its width, and $\phi$ is the rotation of the major axis from horizontal. The data were fit with a Levenberg-Marquardt algorithm and least squares statistic, with 3$\sigma$ outliers rejected iteratively over two passes. The corresponding FWHM $\theta$ is calculated as:
\begin{equation}\label{mofFWHM}
\theta =2\alpha\sqrt{2^{1/\beta}-1}
\end{equation}
Note that unless otherwise indicated, the Moffat minor FWHM will be used to describe PSF size in further analyses in order to remove the contribution of telescope jitter apparent in the AO-off images.

\begin{figure}[t!]
    \centering
    \includegraphics[width=0.483\textwidth]{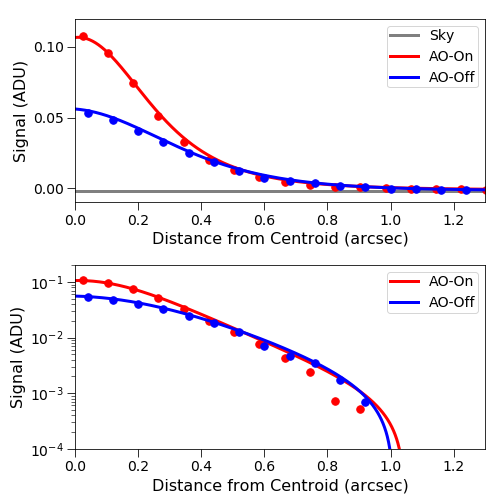}
    \caption{Radial profiles along the minor axis of PSFs in AO-off (blue) and AO-on (red) images, shown with linear (top) and logarithmic (bottom) y-axes. In each case, the image data is represented by circular points and the best-fit model is shown as a solid line. Both PSFs came from the median sized star in stacked images from 2017-05-20 UT and are representative of most of our data.}
    \label{fig:PSFprof}
\end{figure}

\begin{figure}[t!]
    \centering
    \includegraphics[width=0.45\textwidth]{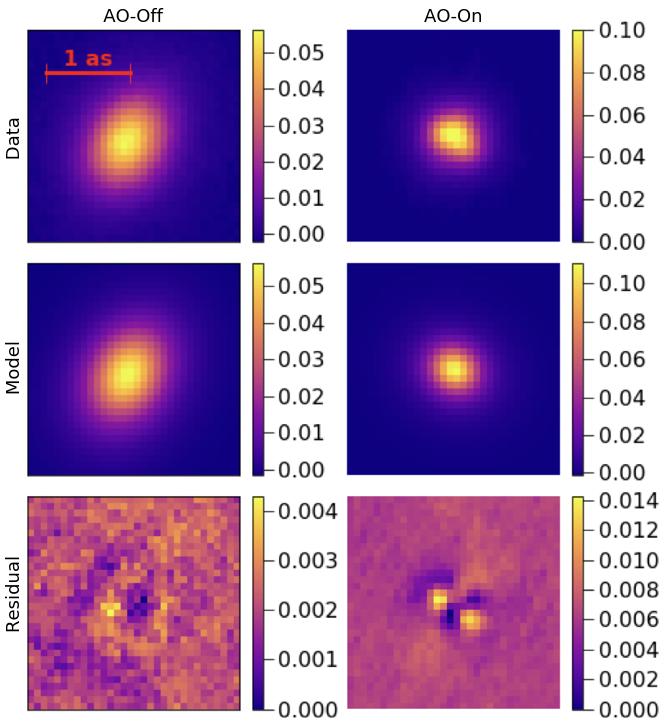}
    \caption{Best-fit PSF model composed of a single component, elliptical Moffat for the AO-off ({\em left column}) and AO-on ({\em right column}) data. The original image ({\em top}), the best fit model ({\em middle}), and the  residuals ({\em bottom)} are shown over the fitting box of 2.56\arcsec by 2.56\arcsec. Colorbar units are in ADU. Image samples were taken from stacked images on 2017-05-20 UT, a night with typical seeing conditions and performance. Note that the color scale is  different between AO-off and AO-on to highlight PSF structure, so the differences in SNR between AO-off and AO-on are not represented here.}
    \label{fig:PSFs}    
\end{figure}

The quality of the model fit is illustrated in Figure \ref{fig:PSFprof}, which shows a radial profile of a AO-off and AO-on PSF and the best fit Moffat profile along the minor axis. There are small discrepancies between the observed and model PSF, particularly in the 2D residuals (Figure $\ref{fig:PSFs}$). The distinct high-order structure evident in the residuals is likely due to the Moffat fitting with a single exponent in the presence of jitter, since the PSF profile follows a different power law along the narrow axis than the elongated one. Static aberrations are calibrated out by optimizing the image quality on artificial sources and recording the centroid offsets on the wavefront sensors, so they are likely not the cause of this residual structure in the PSF. However, the telescope environment may introduce differential aberrations on the various wavefront sensors or some poorly controlled mirror modes could introduce such a pattern once the loop is closed and noise is injected into the system. Table \ref{table:res} compares the result of the Moffat fit, specifically, the RMS of the minor and major FWHM to the previously described metrics.

\begin{figure*}[htb!]
    \centering    \includegraphics[width=\textwidth]{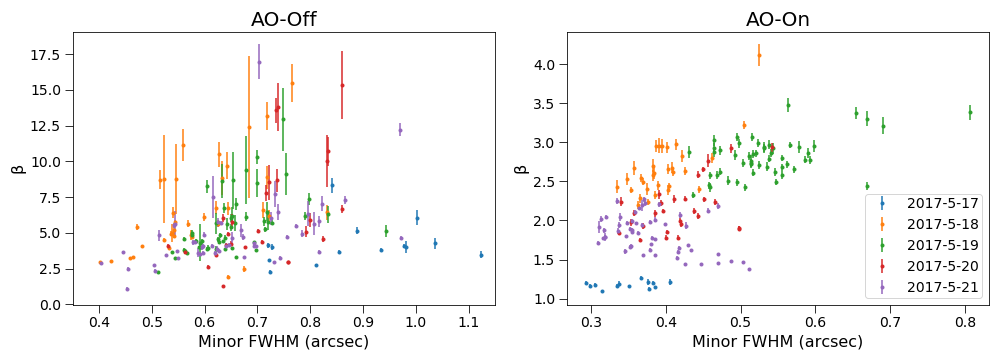}
    \caption{Comparisons of the slope parameter $\beta$ in the Moffat profile to the minor FWHM of the same source. The data show each night of Run 3 in a different color, with each data point representing the median value PSF for a single frame. Points without error bars have uncertainties smaller than the marker size.  \textit{Left:} AO-off (median $\beta$=4.82). \textit{Right: }AO-on (median $\beta$=2.27). }
    \label{fig:beta}
\end{figure*}

The seeing-limited AO-off images are well described by a Moffat profile with a median $\beta\simeq4.8$, which is slightly above the $\beta$ of roughly 3-4 typically measured in astronomical images \citep{saglia1993effects}. Additionally, they show relatively little dependence on conditions or FWHM, as expected (Figure \ref{fig:beta}). For the AO-on images, there is a correlation between $\beta$ and FWHM.

\subsection{\label{sec:wave_calib}Wavelength Conversion}
As previously stated in section \ref{sec:obs}, the data comprising our observations were taken through three different filters (details in Table \ref{tab:log}). For ease of comparison, the MASS/DIMM data (which are observed at 500 nm) are converted to the wavelength of the corresponding `imaka science images. The telemetry data have been similarly converted. The data are converted in the following way: 

For a given wavelength $\lambda$, the Fried parameter describing the amount of wavefront distortion due to atmospheric turbulence scales as:
\begin{equation}\label{proportion}r_o \propto \lambda^{6/5}\end{equation}
To scale a given seeing measurement $\theta_{obs}$ observed at a wavelength, $\lambda_{obs}$, we assume an infinite outer scale and that the seeing measurement is proportional to the ratio of the imaging wavelength and the Fried parameter, which gives  
\begin{equation}\label{seeing}\theta \propto \frac{\lambda}{r_o} \propto \lambda^{-1/5}\end{equation} 
\begin{equation}\label{calibration}\theta_{conv} = \theta_{\mathrm{obs}} \cdot \left(\frac{\lambda_{\mathrm{obs}}}{\lambda_{conv}} \right)^{1/5}
\label{eqn:wave_cal}
\end{equation}
where $\theta_{conv}$ is the seeing scaled to a wavelength $\lambda_{conv{}}$.

\section{\label{sec:results}Results}
\subsection{\label{sec:res_glao_perf}GLAO Performance}

\begin{figure}[b!]
    \centering
    \includegraphics[width=0.5\textwidth]{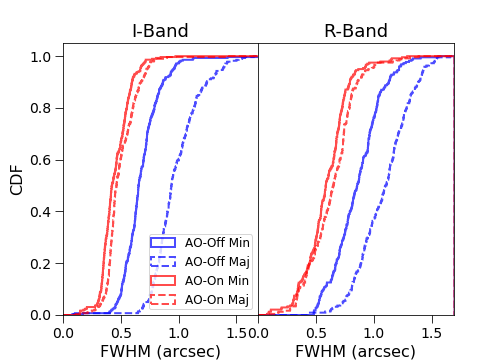}
    \caption{A comparison of the FWHM in AO-off ({\em blue}) and AO-on ({\em red}) images. Minor FWHM is shown in {\em solid} lines, and major FWHM in {\em dashed} lines. Data are divided between $I$-band {\em left} and $R$-band {\em right}.}
    \label{fig:mofCDF}
\end{figure}

The simplest analysis that can be done to characterize the GLAO performance is to compare the FWHM in AO-off images to that in AO-on images, as shown in Figure $\ref{fig:mofCDF}$. The FWHM along the minor axis decreases from 0.76\arcsec\space in the AO-off case to 0.45\arcsec\space with AO correction in $I$-band, a change of 51\%. Changes in all other PSF size metrics at all observed wavelengths are given in Table \ref{tab:res}, and show roughly consistent improvements. 

\begin{deluxetable}{rrrr}[h]
	\tabletypesize{\footnotesize}
    \tablewidth{20pt}
    \tablecaption{PSF Size Improvement \label{tab:res}} 
	\tablehead{
    	Metric &
     	AO-Off &
     	AO-On &
		Change \\
         & (arcsec) & (arcsec) &
	}
	\startdata
    \hline
    & \textbf{$R$-Band (658 nm)} & & \\
    \hline
    Moffat FWHM (min) & 0.857 $\;\pm\;$ 0.014 & 0.589 $\;\pm\;$ 0.012 & 37\% \\
    Moffat FWHM (maj) & 1.097 $\;\pm\;$ 0.016 & 0.639 $\;\pm\;$ 0.013 & 53\% \\
	Empirical FWHM    & 1.008 $\;\pm\;$ 0.015 & 0.666 $\;\pm\;$ 0.013 & 41\% \\
	NEA Width         & 1.000 $\;\pm\;$ 0.014 & 0.736 $\;\pm\;$ 0.015 & 30\% \\
	50\% EE Diameter  & 1.221 $\;\pm\;$ 0.030 & 0.899 $\;\pm\;$ 0.031 & 30\% \\
	80\% EE Diameter  & 2.101 $\;\pm\;$ 0.026 & 1.790 $\;\pm\;$ 0.032 & 16\% \\
    \hline
    & \textbf{$I$-Band (806 nm)} \\
    \hline
    Moffat FWHM (min) & 0.761 $\;\pm\;$ 0.059 & 0.450 $\;\pm\;$ 0.007 & 51\% \\
    Moffat FWHM (maj) & 1.089 $\;\pm\;$ 0.070 & 0.499 $\;\pm\;$ 0.008 & 74\% \\
	Empirical FWHM    & 0.845 $\;\pm\;$ 0.012 & 0.498 $\;\pm\;$ 0.008 & 52\% \\
	NEA Width         & 0.822 $\;\pm\;$ 0.011 & 0.565 $\;\pm\;$ 0.008 & 37\% \\
	50\% EE Diameter  & 1.255 $\;\pm\;$ 0.020 & 0.985 $\;\pm\;$ 0.023 & 24\% \\
	80\% EE Diameter  & 1.799 $\;\pm\;$ 0.020 & 1.435 $\;\pm\;$ 0.014 & 22\% \\
    \hline
    & \textbf{1000 nm} \\
    \hline
    Moffat FWHM (min) & 0.663 $\;\pm\;$ 0.037 & 0.377 $\;\pm\;$ 0.025 & 55\% \\
    Moffat FWHM (maj) & 0.868 $\;\pm\;$ 0.047 & 0.416 $\;\pm\;$ 0.023 & 70\% \\
	Empirical FWHM    & 0.796 $\;\pm\;$ 0.036 & 0.456 $\;\pm\;$ 0.023 & 54\% \\
	NEA Width         & 0.715 $\;\pm\;$ 0.038 & 0.516 $\;\pm\;$ 0.028 & 32\% \\
	50\% EE Diameter  & 0.849 $\;\pm\;$ 0.020 & 0.682 $\;\pm\;$ 0.027 & 22\% \\
	80\% EE Diameter  & 1.468 $\;\pm\;$ 0.031 & 1.304 $\;\pm\;$ 0.035 & 12\% \\
	\enddata
    \tablecomments{Mean PSF sizes by several different metrics, in the AO-off and AO-on. All values are give in arcseconds. Reported errors are the uncertainty in the mean. The data have been divided between observation filter.}
	\label{table:res}
\end{deluxetable}

\begin{figure}[t!]
	\centering
    \includegraphics[width=0.45\textwidth]{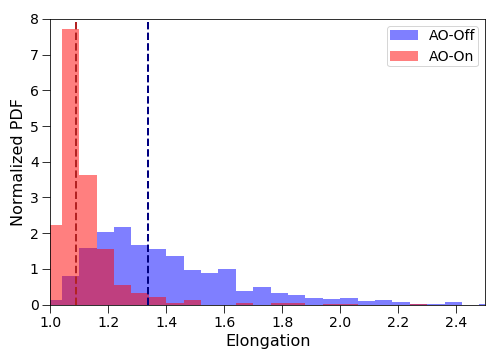}
    \caption{A comparison of the elongation (derived from the Moffat fit and defined by Equation \ref{eqn:elongation}) of PSFs in AO-off ({\em blue}) and AO-on ({\em red}) images. The median value of each distribution is indicated by the vertical {\em dashed line} in the corresponding color: 1.34 for AO-off and 1.08 in AO-on. The elongation in the closed-loop images is consistent with the 8\% plate scale variation in the optical design.} 
	\label{fig:elon}
\end{figure}

Though already showing a fair amount of correction, this comparison does not capture the full impact of the instrument. As was visible in Figure \ref{fig:PSFs}, a change occurs not only in the PSF size, but in its shape as well.  Static and dynamic aberrations within the telescope and instrument are also corrected by the GLAO system.  As noted above the static aberrations are removed in the AO-off images by applying average voltages to the deformable mirror during AO-off images.  The correction of the PSF elongation, caused predominately by telescope jitter, is readily seen in the images in Figure \ref{fig:PSFs}. AO-off images show elongation in one direction by a factor of $\sim$1.3. We attribute this elongation to telescope jitter, as it is generally oriented along either the East-West direction (i.e. RA drive of the telescope). The elongation could be caused by a combination of astigmatism and focus if the deformable mirror voltages used for AO-off images were not averaged for a sufficiently long period.  However, this effect would give rise to a random direction and amplitude to the image elongation.  Since the direction of the PSF elongation is constant over many hours, its unlikely that a static aberration from the DM shape during AO-off images is not the dominant source of elongation. Regardless, elongation is well corrected by the AO system and is nearly nonexistent in the AO-on images. This stark difference can be seen in Figure \ref{fig:elon}, which shows the elongation parameter:

\begin{equation}\label{eqn:elongation}
E=\frac{\theta_{maj}}{\theta_{min}}
\end{equation}
The median value of the elongation decreases from 1.34 to 1.08. This residual elongation of 8\% is consistent with the plate scale variation in the `imaka design, which is roughly 8\% in the East-West direction. Even more, the AO-on elongation values demonstrate substantially less spread than the AO-off values, with standard deviations of 0.28 and 0.35, respectively. 

 In addition to changes in FWHM, we can use our other PSF metrics to quantify `imaka's gains. As seen in Table \ref{tab:res}, the AO-on/AO-off change in the 50\% EE diameter is greater than the same change in  the 80\% diameter (e.g., in $R$-band, the improvements in 50\% EE and 80\% EE are respectively 24\% and 16\%). The relatively low change in the 80\% diameter is indicative of the less substantial correction the system makes to the extended halo. The radius in which any correction of the PSF is possible for a given AO system is set by the number of actuators in the DM. `imaka's current DM has 36 actuators; increasing this number could potentially the radius of its AO-correction, and subsequently increase this amount of correction for the higher encircled energies.
 
\begin{figure}[h!]
	\centering
    \includegraphics[width=0.5\textwidth]{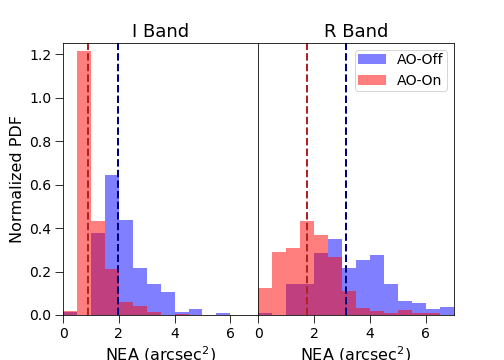}
    \caption{The change in noise equivalent area between AO-off (blue) and AO-on (red) images across all nights, shown as normalized probability distribution functions. Median values of both distribution are designated by vertical dashed lines. The data are divided into $I$-band {\em left} and $R$-band {\em right}.}
	\label{fig:NEA}
\end{figure}

The improvement in our final metric, noise-equivalent area (NEA), is shown in Figure \ref{fig:NEA}. Note that unlike other metrics and the representation of NEA previously, this figure reports NEA as an area; this representation is useful in addition to the NEA width because photometric precision scales with NEA, while astrometric precision scales with NEA$^{1/2}$. The NEA decreases from 1.96 square arcseconds to 0.89 in $I$-band, a change of over a factor of two.     

\begin{figure*}[!]
 	\includegraphics[width=\textwidth,height=6cm]{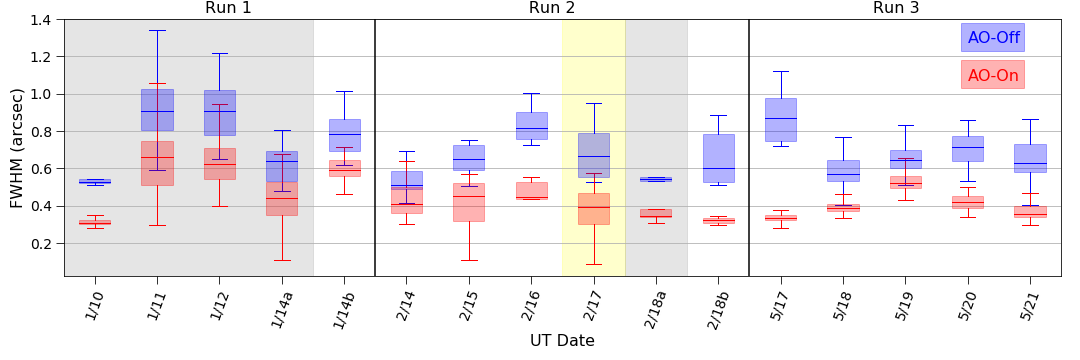}
 	\caption{A comparison of the PSF stability between AO-off ({\em blue}) and AO-on ({\em red}) across all nights. For each data point, the median value of the empirical FWHM is represented by the center line, with a box surrounding it and spanning the second and third quartiles of the data. The full range of data is shown by the extended lines. Shaded regions refer to the observation wavelength: $R$-band in {\em gray}, 1000 nm in {\em yellow}, and $I$-band is unshaded.}. 
 	\label{fig:stability}
\end{figure*}

Besides improvements in individual PSFs, AO-on images demonstrate higher stability throughout each night of observing. The spread of minor FWHM for each night in both AO-off and AO-on images can be seen in Figure \textbf{$\ref{fig:stability}$}. Regardless of seeing conditions for a given night, there is noticeably more variation in the size of AO-off PSFs than those of the AO-on images.  This effect is particularly apparent when the free-atmosphere is weak.  This is well illustrated in the May run, where the night-averaged standard deviation of FWHM decreases from 0\farcs59 to 0\farcs05, a change of more than an order of magnitude.  In contrast, when the total seeing is dominated by poor free-atmosphere seeing (e.g. Run 1), the PSF variability even in GLAO mode, is more pronounced.   This trend can also be seen with more detail in Figure \textbf{$\ref{fig:Nightly}$}, where the same data are shown with MASS/DIMM data over time for each night. Generally, even when the ground layer seeing has more variation over the course of the night than the free atmosphere, the AO-on FWHM remains more stable than in the AO-off case.

\subsection{\label{sec:res_vs_seeing}Comparisons to Seeing}

\begin{figure}[!]
    \centering
    \includegraphics[width=0.45\textwidth]{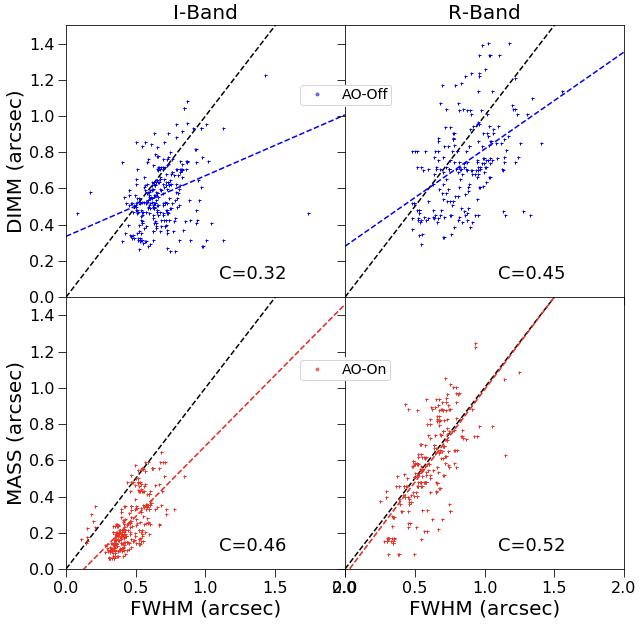}
    \caption{Correlations between focal plane data and MASS/DIMM.  {\em Top:} correlation between AO-off images and DIMM seeing, matched by observation time. {\em Bottom:} AO-on images and MASS. In each panel, data points are shown with a line of best fit in the same color, with the corresponding correlation coefficient C reported. A dashed black line in each panel shows a 1:1 correlation. The AO-on/AO-off data are separated by filter, with $I$-band on the left and $R$-band on the right. The MASS/DIMM are converted to match the wavelength of the `imaka image data using equation \ref{eqn:wave_cal}. Note that the I-band and R-band data sets were taken on different sets of nights with different seeing conditions.}
    \label{fig:corrs}
\end{figure}

The comparison between AO-off and AO-on images already begins to characterize the gains from GLAO. However, a more objective indicator of success would be to compare the GLAO PSFs to the seeing above the ground layer using MASS/DIMM measurements described in \S\ref{sec:obs_mass_dimm}.  First, we compare the FWHM of the focal plane images to the contemporaneous MASS/DIMM seeing  in Figure \ref{fig:corrs}, where the top and bottom panels show the correlations between AO-off images and DIMM, and AO-on images and MASS, respectively.  

Assuming accurate measurements by the MASS/DIMM instruments and no error in our correction, the DIMM should show some correspondence with the AO-off image FWHM and similarly the MASS should correspond to the AO-on data. To examine this relationship, we looked at the correlation between the focal plane data, telemetry, and MASS/DIMM seeing. Across all nights, we calculated the correlation coefficient for six combinations: three in the total atmosphere regime (AO-off vs. free atmosphere telemetry; AO-off vs. DIMM; integrated seeing telemetry vs. DIMM) and the free atmosphere regime (AO-on vs. integrated seeing telemetry; AO-on vs. MASS; free atmosphere seeing telemetry vs. MASS). In the case of AO-on images, we see good correlation with the MASS, with a correlation coefficient of 0.46 in $I$-band and 0.52 in $R$-band. These correlations are shown in Figure \ref{fig:corrs}. The image data is similarly well correlated with `imaka's telemetry in $I$ and $R$ band, though some discrepancies occur at 1000nm. We note that the amount of data taken in this filter is small, so the lack of correlations here may not be significant. A full list of correlation coefficients is presented in Table \ref{tab:corr}.

Although the AO-on images are correlated with the MASS seeing, there is a distinct offset between their cumulative distributions with the MASS seeing being smaller than the AO-on FWHM.  One limiting factor in the agreement is the apparent floor in the AO-on images at approximately 0\farcs3.  Across all analyses, our GLAO PSF sizes generally don't go below this minimum even when the seeing predicted by the MASS does drop below 0\farcs3. This limitation is likely due to systematic effects (e.g., static aberrations, internal seeing), that the low order system (due to a relative lack of DM actuators and the low bandwidth of our wavefront sensor cameras) cannot correct.

\begin{deluxetable}{lccc}[h]
	\tabletypesize{\footnotesize}
    \tablewidth{0.5\textwidth}
    \tablecaption{Comparison of Image Quality with Seeing Estimates \label{tab:see}} 
	\tablehead{
	 & Total Atm. & Free Atm. & Change \\
     & or AO-Off & or AO-on &
	}
   	\startdata
     & (arcsec) & (arcsec)  & \\
    \hline
    & \textbf{$R$-Band (658nm)} & & \\
    \hline
	`Imaka Images & 0.857 $\pm$ 0.014 & 0.589 $\pm$ 0.012 & 37\% \\
	`Imaka Telemetry & 0.854 $\pm$ 0.014 & 0.665 $\pm$ 0.012 & 25\% \\
	MASS/DIMM & 0.728 $\pm$ 0.011 & 0.616 $\pm$ 0.013 & 17\% \\
    \hline
    & \textbf{$I$-Band (806nm)} & & \\
    \hline
	`Imaka Images & 0.761 $\pm$ 0.059 & 0.45 $\pm$ 0.007 & 51\% \\
	`Imaka Telemetry & 0.572 $\pm$ 0.012 & 0.401 $\pm$ 0.009 & 35\% \\
	MASS/DIMM & 0.567 $\pm$ 0.007 & 0.253 $\pm$ 0.006 & 76\% \\
    \hline
    & \textbf{1000nm} & & \\
    \hline
	`Imaka Images & 0.663 $\pm$ 0.037 & 0.377 $\pm$ 0.025 & 55\% \\
	`Imaka Telemetry & 0.563 $\pm$ 0.023 & 0.453 $\pm$ 0.023 & 22\% \\
	MASS/DIMM & 0.445 $\pm$ 0.011 & 0.297 $\pm$ 0.021 & 40\% \\
	\enddata
    \tablecomments{Mean PSF sizes by three different measurements, observing the total integrated atmosphere (from top to bottom: AO-off images, AO-off telemetry, MASS integrals) and the free atmosphere (AO-on images, AO-on telemetry, DIMM integrals). All values are give in arcseconds. Reported uncertainties are the error on the mean. The data are separated by observation wavelength of the `imaka images, but all data are shown at their original wavelengths. }
    \label{table:see}
\end{deluxetable}

\begin{figure}[h]
  \includegraphics[width=0.5\textwidth]{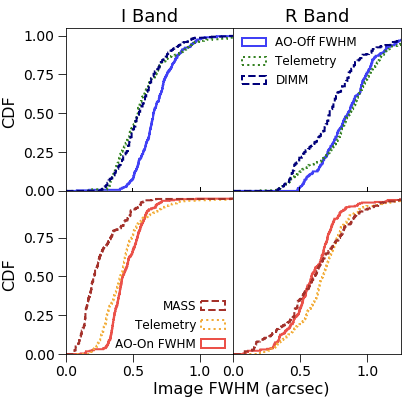}
  \caption{Cumulative distribution functions comparing image plane, telemetry, and MASS/DIMM data. The top panels show values associated with the total atmosphere: telemetry and DIMM seeing, and are compared to the AO-off image plane PSF size. The bottom panels show the free atmosphere case: the telemetry case is for all turbulence above $\sim 600 m$, along with MASS and AO-on image plane PSF size. The AO-on/AO-off data are separated by filter, with $I$-band on the left and $R$-band on the right. The MASS/DIMM and telemetry, taken from nights that corresponded to each set of observation data, are converted to match the wavelength of the `imaka image data using equation \ref{eqn:wave_cal}. Note that the I-band and R-band data sets were taken on different sets of nights with different seeing conditions. } 
  \label{fig:CDFs}
\end{figure}

As mentioned in \S3.3, comparisons to MASS/DIMM are also limited by the fact that they are not observing along the same lines of sight through the atmosphere. To compensate for this, we compare the image quality to our estimates of the integrated and free-atmosphere seeing derived from our telemetry measurements. These have the advantage that these data are synchronized with the images.  Figure \textbf{$\ref{fig:CDFs}$} compares the science image PSF FWHM to both the telemetry measured by `imaka and to the seeing measured by the MASS/DIMM. Median values and standard deviations are given for all these data sets in Table \ref{tab:see}. 

Figure \textbf{$\ref{fig:CDFs}$} shows the cumulative distributions for the image FWHM and the seeing from the MASS/DIMM and the imaka telemetry.  There is a mix of agreement and disagreement between the distributions of seeing and image FWHM.  During the nights when we observed in $I$-band, both the AO-on and AO-off images FWHM distributions were shifted to values larger than the I-band converted MASS/DIMM seeing and telemetry derived seeing. In these cases the distribution of MASS seeing values is significantly lower than either the telemetry derived free-atmosphere seeing or the AO-on image FWHM.  On the other hand, on nights when we observed in $R$-band, the AO-off image FWHMs are smaller than the telemetry seeing but worse than the MASS/DIMM seeing but the AO-on image FWHMs are comparable or better than the MASS or telemetry derived free-atmosphere seeing.  

Some of this difference could come from the fact that the $I$-band and $R$-band nights had quite different seeing conditions.  During most of the nights when we observed in $I$-band, the free-atmosphere seeing was weak and both the MASS and DIMM seeing are better than their corresponding image FWHMs. During the the $R$-band nights the total seeing was dominated by large free-atmosphere seeing. Here, the MASS seeing CDF and AO-on image CDF are similar for the better free-atmosphere seeing but skewed to values larger than the AO-on image FWHM CDF in the poorer free-atmosphere seeing conditions. Together, the $I$-band and $R$-band AO-on data suggest that we have an instrumental floor which prevents GLAO images of better than about 0.3 arcseconds.  Alternatively, the data may be indicating that the MASS underestimates the free-atmosphere seeing under good seeing conditions, as seen in \cite{lombardi2015using}. 

For the telemetry data the shapes of the distribution functions are similar in all cases to the distributions for the image FWHM, but with fixed offsets between them. One possibility is that the effective wavelength of the wavefront sensors changes between the observed fields.  All of the $R$-band data were taken on the Pleiades field where the guide stars are generally bluer than the guide stars in the other fields.  Since we assume a wavelength of 0.7 microns for the WFSs, we may be overestimating our seeing estimates in Pleiades field (e.g. $R$-band data).  Alternative explanations could be an effect that happens on longer time scales, such that it is seen in the 30-45 second science camera exposures but not the 15-second telemetry exposures. Additionally, if the dome seeing consists of mostly high frequencies (which the AO system can not correct for), these residual wavefront aberrations could contribute to a broadening of the wings of the PSF, in turn slightly enlarging the FWHM.

\begin{deluxetable}{lcc}[]
	\tabletypesize{\footnotesize}
    \tablewidth{15pt}
    \tablecaption{Correlations Coefficients between Data and Seeing \label{tab:corr}} 
	\tablehead{
     & Total Atmosphere & Free Atmosphere \\
     & (AO-off/DIMM) & (AO-on/MASS)
	}
   	\startdata
    & \textbf{$R$-Band (658nm)} \\
    \hline
    Focal Plane vs. Telemetry & 0.68 & 0.70 \\
    Focal Plane vs. MASS/DIMM & 0.45 & 0.52 \\
    Telemetry vs. MASS/DIMM & 0.35 & 0.47 \\
    \hline
    & \textbf{$I$-Band (806nm)} \\
    \hline
    Focal Plane vs. Telemetry & 0.33 & 0.36 \\
    Focal Plane vs. MASS/DIMM & 0.32 & 0.46 \\
    Telemetry vs. MASS/DIMM & -0.10 & 0.23 \\
    \hline
    & \textbf{1000nm} \\
    \hline
    Focal Plane vs. Telemetry & 0.15 & 0.43 \\
    Focal Plane vs. MASS/DIMM & -0.02 & 0.14 \\ 
    Telemetry vs. MASS/DIMM & -0.23 & 0.27 \\
	\enddata
    \tablecomments{Correlation coefficients for `imaka image data, `imaka telemetry, and MASS/DIMM seeing in both the total and free atmosphere cases. Telemetry and MASS/DIMM data have been converted to the matching observation wavelength with equation \ref{eqn:wave_cal}.}
\end{deluxetable}

\subsection{\label{sec:res_vs_wave}Wavelength Dependence}
\begin{figure}[h]
    \centering
	\includegraphics[width=0.483\textwidth]{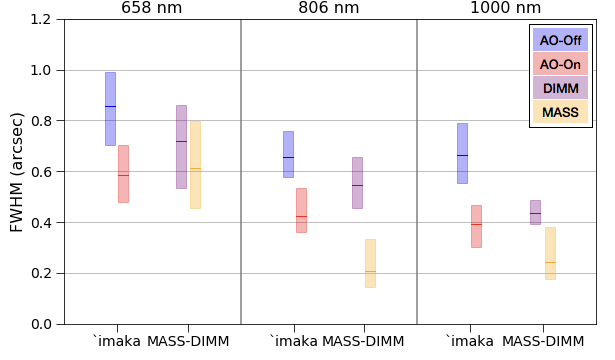}
    \caption{Wavelength dependence of PSF size is shown here as minor Moffat FWHM for AO-off (\textit{blue}) and AO-on (\textit{red}) images for all nights, with no wavelength conversion applied. Each set is shown next to the MASS (\textit{orange}) and DIMM (\textit{purple}) seeing measurements for the same observation times. As the MASS/DIMM data is initially measured at 500 nm, their values are presented here scaled to the wavelength of the corresponding `imaka data points. Each box represents the mean value with a solid line and the second and third quartiles with a shaded region. We note that the $R$-band and $I$-band points are from an average of $\geq$5 nights, while the 1 $\mu$m point is from only one night (refer to Table \ref{tab:log} for observation details).}
      \label{fig:wvl}
\end{figure}

We can examine the wavelength dependence of GLAO correction by plotting the raw FWHM values at different wavelengths, as shown in Figure \textbf{$\ref{fig:wvl}$}. Ideally, we would simply look at how FWHM varies with wavelength. However, because the observations in the different filters were done over different nights and over a wide range of observing conditions, direct comparisons cannot be made from night to night.  For this reason, we present each data set with the MASS/DIMM seeing measurements for the corresponding times, converted from 500 nm to the relevant wavelength. As in previous figures, we see an offset between AO-off and DIMM, as well as between AO-on and MASS. Despite this discrepancy, at each wavelength there is a clear improvement between AO-off and AO-on, sometimes larger than the difference between DIMM and MASS. Though we do not yet have enough data in consistent seeing conditions to conclude anything about the specific dependence of GLAO correction on wavelength, we can already see that `imaka achieves substantial correction at all currently observed wavelengths.

\begin{figure*}
 	\includegraphics[width=\textwidth,height=6cm]{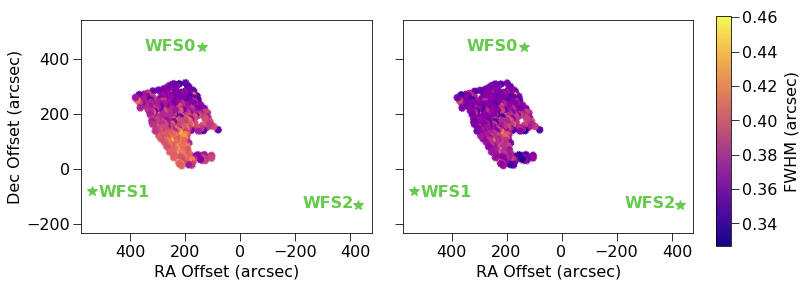}
    \centering
    \caption{The FWHM of individual stars in a stacked image of all AO-on frames from 2017-05-18 UT.  \textit{Left:} The original data, showing the variability of the PSF across the science field. \textit{Right:} The same data after subtraction of the best fit plane, with the median data point from the left (0.40\arcsec) added to all points for comparison purposes. Wavefront sensors' positions relative to the science frame are marked in green and WFS1 is the brightest guide star. The data were taken at $I$-band with no wavelength conversion applied. By visual inspection, the field on the left shows more structure in its variability, with a strong gradient in FWHM increasing to the bottom left. This global pattern in the variability is largely removed in the right panel.  Between the two figures, the range of FWHM decreases from 0.201\arcsec on the left to 0.110\arcsec on the right.}
    \label{fig:Scatter}
    \end{figure*}

\subsection{\label{sec:level2}Field Variability}

From simulations the GLAO PSF is predicted to be quite uniform across the field.  However, as GLAO is deployed over wider and wider fields, it is important to ensure that the instrumental PSF quality does not suffer with increasing field size.  The left panel of Figure \ref{fig:Scatter} shows the FWHM of each star as a function of position in the stacked AO-on image from 2017-05-19 UT (chosen because its seeing conditions were typical of the run), with the location of guide stars marked in green.  To better view any global trends across the field, these data had outliers removed through an iterative sigma clipping routine, rejecting points more discrepant than 3$\sigma$ from the median, with 5 iterations.

The dominant field variation is a gradient in FWHM, with PSFs increasing in size towards the East side of the field. In single guide star AO, the expectation is for the PSF quality to improve closer to the guide star. The trend visible here is not well correlated to the distance to the field center or to any single guide star. The trend is repeated over multiple nights and appears fixed to the sky, as we describe in more detail below.

\begin{deluxetable}{lccrr}[]
	\tabletypesize{\footnotesize}
    \tablewidth{15pt}
    \tablecaption{Nightly Variability Plane Fit for Run 3 			\label{tab:var}} 
	\tablehead{
		Date &
	$\Delta FWHM_{i}$ &
    $\Delta FWHM_{f}$ &
    Direction $\phi$  &
    Gradient \\
    (UT) & (") & (") & ($^{\circ}$) & (arcsec/arcmin)
	}
   	\startdata
    2017-5-17 & 0.128 & 0.032 & $153.4\pm6.0$ & $(16.6\pm0.2)\times10^{-3}$ \\ 
	2017-5-18 & 0.201 & 0.110 & $97.0\pm9.7$ & $(8.8\pm0.1)\times10^{-3}$ \\ 
	2017-5-19 & 0.104 & 0.091 & $95.8\pm9.7$ & $(8.2\pm0.1)\times10^{-3}$ \\ 
    2017-5-20 & 0.115 & 0.086 & $86.0\pm14.5$ & $(5.0\pm0.1)\times10^{-3}$ \\ 
    2017-5-21 & 0.178 & 0.092 & $92.3\pm31.3$ & $(4.2\pm0.1)\times10^{-3}$ \\ 
	\enddata
    \tablecomments{For each night of the May run, the mean range of FWHM values of all stars in AO-on images is given for both before ($\Delta FWHM_{i}$) and after ($\Delta FWHM_{f}$) the removal of a best fit plane. The value $\phi$ represents the direction of the best fit plane (in degrees, East of North with respect to the detector) while the gradient describes the variation in arcseconds per arcminute of the plane.}
    \label{ta:var}
\end{deluxetable}

We used several methods of quantifying the global variation. First, the data shown in the left panel of Figure \ref{fig:Scatter} was fit in three dimensions (x, y, and FWHM) to a plane. The best-fit plane is defined by two quantities. The first, $\phi$, is the position angle of the plane's normal vector projected onto the focal plane and measured East of North in degrees.
The second is the gradient, which measures the spatial change of FWHM in arcseconds per arcminute. The right panel of Figure \ref{fig:Scatter} shows the residual FWHM after subtraction of the best-fit plane for the stacked image on 2017-05-19 UT, with an offset of 0\farcs44 added for the purpose of comparison. In addition to these parameters, we calculated the range of FWHM values before and after the plane removal ($\Delta FWHM_{i}$ and $\Delta FWHM_{f}$).

As seen in Figure \ref{fig:Scatter}, though some faint, high-order structure appears to remain, the variation decreases from the left to the right panels substantially.  Averaged over the five nights of the May run, the range decreases from 0.145\arcsec to 0.082\arcsec.  Finally, the average variation of FWHM across the best-fit planes in the direction of greatest change is 0\farcs0086 per arcminute of the field.

To justify the removal of the fitted plane, the variability would have to be attributed to an instrumental effect, such as a tilt in the CCD relative to the incoming beam. To check whether the planar variability could be due to an atmospheric effect, similar plane fits were generated for each individual frame on the same night. On a frame-by-frame basis for all images taken 2017-05-18 UT, neither the direction nor the gradient showed significant correlation to time, ground layer seeing, wind speed or direction, or zenith angle of the telescope (a potential cause of flexure in the instrument). Based on the lack of any trends in these analyses, we conclude that the field variability of the FWHM is dependent on an instrumental effect such as misalignment, which will be addressed in the future.

Further evidence for an instrumental origin of the PSF field-variability comes from the stability of the plane over multiple nights. The stacked AO-on images from each night of the 2017 May run were fit with a plane in the same manner as above and results are listed in Table \ref{tab:var}. The plane parameters $\phi$ and gradient are given, along with $\Delta FWHM_{i}$ and $\Delta FWHM_{f}$. The uncertainties in angles were generated using a full sample bootstrap with replacement routine in the fit. In order to generalize this effect, we report 4.5\% as the typical variation in FWHM across the field, calculated as the mean of each night's percent change from the median to the minimum point. As such, we can tentatively apply a correction of 4.5\% to any focal plane PSF size measurements due to field variability.

The variation in the data of Table \ref{tab:var} may be explained by adjustments made to the instrument between nights.  For example, the science camera was removed and rotated by 45$^{o}$ between 2017-05-17 UT and 2017-05-18 UT, corresponding to the large change in the direction of variation.  Also, between 2017-19-20 and 2017-5-20, the camera was taken off and returned to the same position. The constant direction with different inclination suggests that  after remounting the camera, we modified the camera tilt slightly. Though not all the changes between nights can be explained with camera changes, these data do present us with potential experiments we can conduct in the future in order to understand the source of this aberration.

It should be noted that the entirety of this variability analysis comprises of only AO-on images. The gradient apparent in these images is not seen in AO-off frames, as the image quality is too poor to detect the small change in the FWHM over the field of view. Any attempts at fitting the variability had uncertainties too large to provide valuable insight, so they were omitted from this analysis.

\section{\label{sec:discussion}Discussion}

The results of our three commissioning runs already show the value of `imaka and GLAO and the science cases it makes possible. From the size of the field alone, the potential to obtain sharp images of extended sources or large, crowded fields is apparent. Across all nights, our science images display an improvement in FWHM of up to a factor of 1.8 , depending on the metric used (summarized in Table \ref{tab:res}). The median ($I$-band) minor FWHM across all nights of 0.45\arcsec$\pm$0.03\arcsec is consistent with Monte-Carlo simulations for `imaka, which predicted FWHM in the same band of approximately 0.45\arcsec-0.50\arcsec within the central 6.5\arcmin \citep{2014SPIE.9148E..1KC}. The range in improvement comes from the difference in how each metric quantifies the shape of the PSF, and can be most simply summarized by looking at the change in minor and major FWHM of the Moffat profile (0.72\arcsec to 0.49\arcsec and 1.00\arcsec to 0.52\arcsec, respectively). The difference along the two axes in AO-off images are indicative of significant elongation, likely due to telescope jitter, which is almost entirely removed in AO-on images, where the median ratio of major to minor FWHM decreases from 1.34 to 1.08. This highlights the value of a GLAO system in cleaning up telescope and instrumental artifacts that would otherwise reduce image quality in ``seeing-limited'' data.

As an alternative metric, the radius of 50\% encircled energy decreased from 1.26\arcsec to 0.99\arcsec in $I$-band. The decrease of 50\% EE radius by a factor of 1.3 corresponds to increasing the depth of observable stars by approximately 0.9 magnitudes at a given exposure time, yielding larger samples of stars in a given field.  This is particularly advantageous for high-precision astrometry and photometry, where the precision of a measurement scales with the square root of the number of reference stars used.

Besides the increased number of observable objects, the improvement in individual PSFs allows for other opportunities. Photometric precision, for example, scales with NEA, which `imaka has improved from 1.96 to 0.89 square arcseconds ($I$-band), a change of more than a factor of 2.  Astrometric precision is also improved by a more concentrated PSF, scaling with the square root of NEA (or what we term the NEA width).

In addition to improving individual PSFs, AO-on images from `imaka demonstrate improved stability in the FWHM over time and through varying seeing conditions. This effect is most clearly seen in the five nights of the May run, where the nightly standard deviation of FWHM went down from 0\farcs59 in AO-off images to 0\farcs05 in AO-on images, a change of nearly an order of magnitude. These improvements greatly aid spectroscopic observation. Besides the fact that increased encircled energy yields a higher SNR for a given exposure time, the stability of the PSF size allows for optimized slit width selection. This optimization in turn allows for maximum spectral resolution in an observing run: Assume, for example, the ratio of maximum FWHM in a night in with AO-off and AO-on is `imaka's average of 1.6, as previously reported. A decrease in  the slit width by this amount would improve the spectral resolution by nearly a factor of 1.6.  In addition, since the GLAO PSF is more stable spatially and temporally, there is the possibility to decrease the slit width further.  

Furthermore, `imaka has demonstrated an amount of correction that could be the difference between an unusable night of observing and a usable night. This is a major benefit of GLAO observing; \citet{andersen2006performance} demonstrated in simulations that at the Cerro Pachon site in Chile, a GLAO system would improve an observatory's efficiency by up to 40\% by increasing the percentage of the `best seeing-limited image quality nights' from 20\% to about 70\%. As is visible in Figure \textbf{$\ref{fig:Nightly}$}, sporadic changes in seeing are minimized by `imaka, in addition to the seeing being significantly decreased. Eliminating nights unusable because of bad seeing would increase the science output of any facility, an advantage for all types of observing.

Considering the fact that `imaka has thus far only been run in natural guide star (NGS) mode, its gains are already comparable to GLAO systems using laser guide stars (LGS). For example, the previously mentioned ARGOS instrument and the GLAO system on the MMT telescope both achieved improvement in FWHM of roughly a factor of two in LGS mode \citep{LBT_glao_2016, Hart2010}. We have come close to this improvement in larger fields while still using natural guide stars, which, for GLAO, inject noise from atmospheric profiles unrelated to the corrected field. The initial success of `imaka compared to such instruments makes the prospect of adding laser guide star capability in the future particularly promising.

In that vein, `imaka was constructed on a low budget, as it is meant primarily as a demonstrator for GLAO on Maunakea.  As such, it exhibits limitations that the high-quality optics and system engineering commonly used in large-telescope instrumentation projects could easily address.

Despite the clear gains `imaka has achieved, there are distinct differences between the distributions of FWHM we observe in our science images and the seeing (both integrated and free-atmosphere) estimated from the `imaka wavefront sensors and Maunakea MASS/DIMM. Though the median MASS measurement of free atmosphere seeing was 0.25\arcsec across all nights of observing in $I$-band, the image quality of AO-on images those nights was reduced to only 0.45\arcsec, while `imaka's telemetry measured 0.40\arcsec seeing for what is ostensibly the same atmospheric profile experienced by the science camera. Despite this offset, the focal plane images are typically correlated to the telemetry and MASS/DIMM. The offset in the cumulative probability distribution (Figure \ref{fig:CDFs}) between the telemetry and the focal plane images indicates an instrumental error; it is nearly constant with image FWHM, suggesting that it is not a fixed non-common path aberration since its contribution to the image degradation would decrease as the seeing degrades.  There is some indication that the tomographic reconstruction error (e.g. our ability to separate ground-layer turbulence from high-altitude turbulence) may be larger than expected but this is left to a later paper to quantify.  There could also be a floor in the `imaka image quality of $FWHM_{min} \sim 0.3\arcsec$ that the AO-on images cannot go below.  Potential causes for this limitation will also be explored in relation to `imaka's error budget in a later paper. 

The initial characterization of AO-on images in relation to wavelength and field position in this paper reveals that we have not yet reached the atmospheric limits with 'imaka and we may be able to improve instrument calibration and performance in the future. Though very preliminary, we can already see a weak trend in FWHM as a function of observation wavelength. Because of the limited amount of multi-wavelength data and constraints on seeing conditions, it is currently difficult to conclude more about the wavelength dependence. However, we plan to extend this analysis in the future by observing both at shorter and longer wavelengths over a wider range of seeing conditions.  As AO correction is substantially more difficult to achieve with shorter wavelengths, we will test how far this limitation can be pushed. 

We will also test larger fields in order to better understand the variability of the GLAO correction across images. Initial analysis of the five nights of the May run show an average range of 0\farcs145. This variation is  reduced to a range of 0\farcs082 when the planar structure is removed. The plane itself varies on average by 0.0082 arcseconds per arcminute across the field, which corresponds to a typical variation (percent change between minimum and median) of 4.5\%. We will continue to examine this effect in future runs, both with larger fields and different asterisms. The data presented in this paper represents one configuration of wavefront sensors and control matrices per night on relatively small science fields; in reality, we collected data with a range of asterisms, which we are currently analyzing in more detail.

This paper has only looked at a narrow range of commissioning results. It serves as a complement to (Chun, in prep), the main results paper of `imaka's commissioning and will be followed by additional papers specifically on `imaka's PSF and `imaka's photometric and astrometric precision. We will also be continuing regular `imaka runs with new experiments. In order to better address some of the unanswered questions in performance thus far, one new component of the coming runs will be the use of larger science cameras, with plans for an $11\arcmin$ optical camera and a $7\arcmin$ infrared camera. This will allow us to further examine the GLAO correction over large areas and at longer wavelengths.

Collectively, the results of `imaka demonstrate the potential for a GLAO system on telescopes on Maunakea.  The improvements we see in image quality across large fields could yield a significant boost in sensitivity for multi-object spectrographs, potential for high precision astrometry, and high spatial resolution of large sources, all of which can be valuable to a wide variety of science cases. Though the delivered image quality of `imaka on the UH 2.2m may not be identical to expectations for larger telescopes, our image quality improvement by a factor of 1.4-1.9 is comparable to the gains made by GLAO systems deployed on larger telescopes, such as the factor of 2 improvement in FWHM achieved by ARGOS on each of LBT's 8.4 m telescopes \citep{LBT_glao_2016} and the improvement of 1.5-2 in FWHM on the VLT's 8.2 m UT4 telescope \citep{2017Msngr.168....8A}. Furthermore, many instruments on Keck, such as LRIS and DEIMOS, are impacted by effects (e.g. focus changes, telescope jitter) that we have shown `imaka can correct \citep{faber2003deimos, rockosi2010low}. 

\section{\label{sec:conclusion}Conclusion}

We have demonstrated the potential of GLAO with guide stars distributed over fields of about 16 arcminutes over 15 nights of observing. These initial findings are showing promising results:

Between AO-off and AO-on modes, `imaka's focal plane images demonstrate a decrease in PSF size by up to a factor of 1.8, depending on the metric used. The AO-on PSF also shows an elongation of 1:1.1, which is consistent no elongation given the design variation of the plate scale of 8\% in the North-South and East-West directions. This represents a substantial reduction from the AO-off value of 1:1.4. The median minor FWHM derived from a Moffat profile for AO-on images was 0\farcs56 in $R$-band and 0.43\arcsec in $I$-band, and 0.82\arcsec for AO-off images in $R$-band and 0.72\arcsec in $I$-band (including a 4.5\% correction for field variability). These quantities are typically correlated with `imaka's telemetry measurements of the free-atmosphere and integrated seeing, with correlation coefficients in $R$-band of 0.68 and 0.70 respectively. Despite variation in the seeing over the course of a night, AO-on images demonstrate significantly more temporal stability than AO-off images, with the standard deviation of minor FWHM decreasing from 0.59\arcsec to 0.05\arcsec, averaged across the five nights of the final run. The variation in FWHM across the field is also minimized in AO-on images, with FWHM varying on average by 0.0082\arcsec per arcminute and an average percent variation of 10.3\%. Overall, these early results from 'imaka  suggest that a facility-class ground-layer adaptive optics system on Maunakea would deliver valuable gains over seeing-limited images and spectroscopy.

\section*{Acknowledgements}

`imaka is supported by the National Science Foundation under Grant No. AST-1310706 and by the Mount Cuba Astronomical Foundation. C.B. acknowledges support from the Alfred P. Sloan Foundation. We would like to acknowledge the assistance of the UH 2.2m day crew for assistance with installation and observing setup.  In addition, significant contributions to the design and construction of `imaka were made by Quartus Engineering, RockWest Composites, and Dream Telescopes.

The authors also wish to recognize and acknowledge the very significant cultural role and reverence that the summit of Maunakea has always had within the indigenous Hawaiian community.  We are most fortunate to have the opportunity to conduct observations from this mountain.

{\it Facilities:} UH:2.2m (`imaka)

\bigskip
\bigskip
\bigskip
\bigskip
\bigskip
\bigskip
\bigskip

\bibliographystyle{aasjournal}
\bibliography{ref}

\appendix

\section{Nightly Performance} \label{sec:nightly}

\begin{figure*}[h]
  \includegraphics[width=0.95\textwidth]{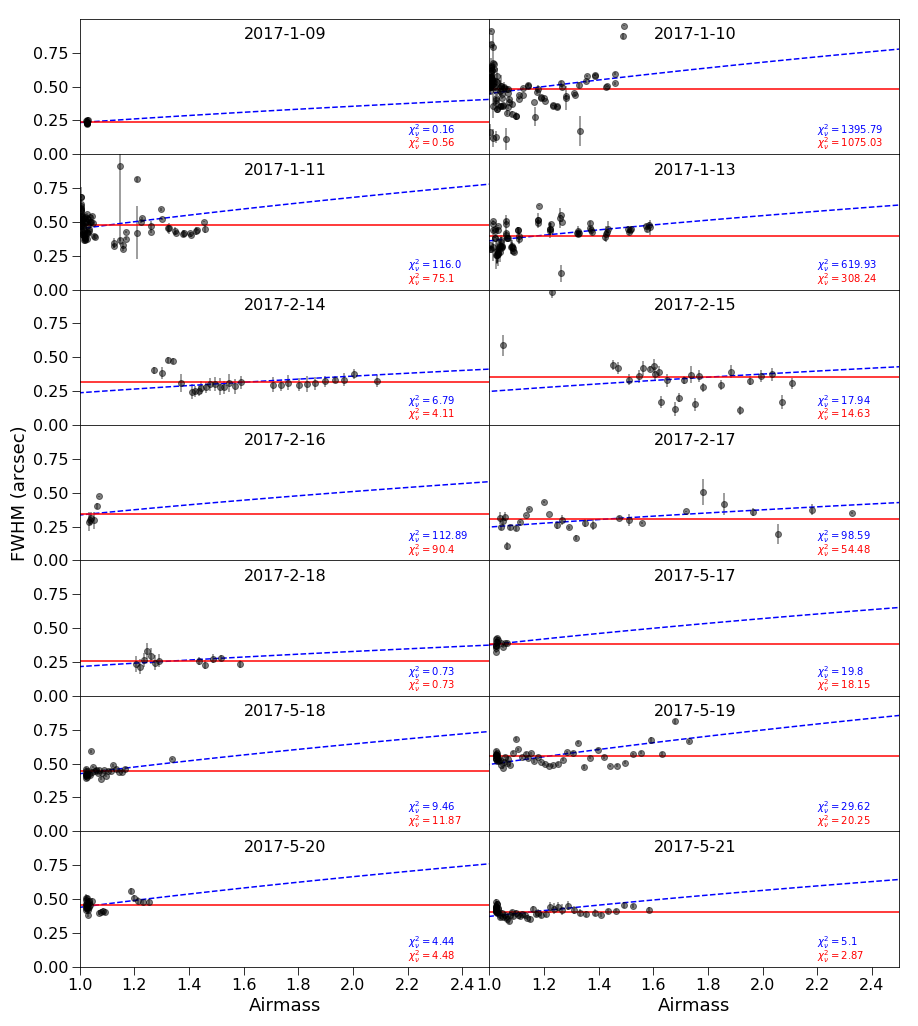}
  \caption{PSF FWHM versus airmass for each night.  Each data point (black) represents the median value of all sources in a frame (AO-on minor FWHM). The data were fit to two models: a constant FWHM (red solid line) and a 3/5 power law (blue dashed line).  The corresponding reduced$-\chi^2$ for both fits are reported in matching colors. } 
  \label{fig:airmass}
\end{figure*}

\begin{figure*}[h]
  \includegraphics[width=0.95\textwidth]{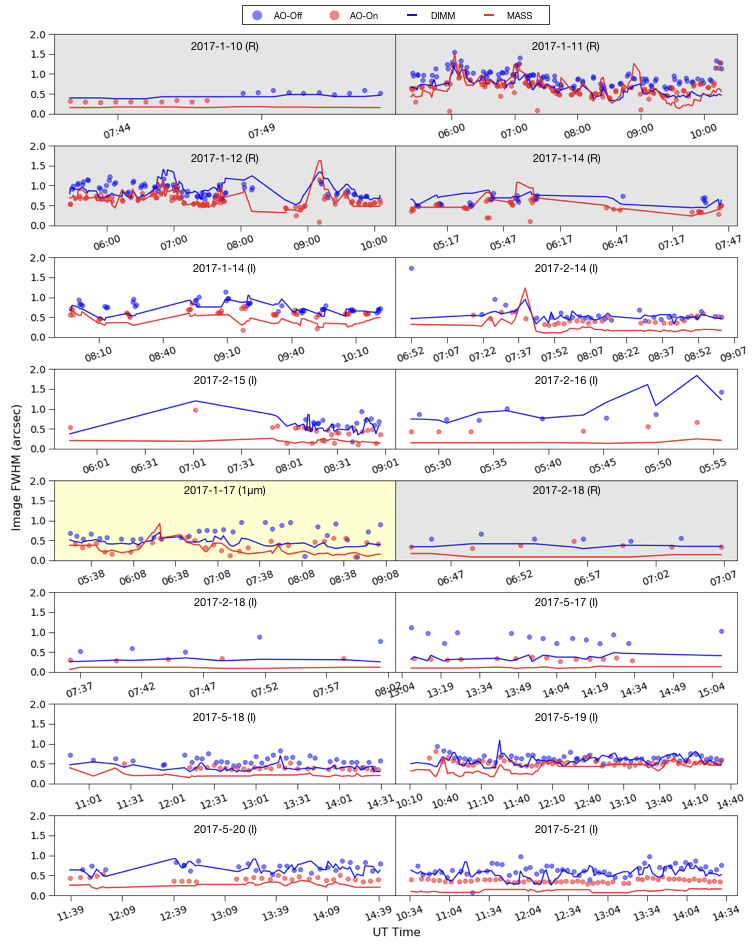}
  \caption{Nightly performance summaries.  In each plot, AO-off and AO-on image FWHM (blue and red dots, respectively) are compared over time with the MASS/DIMM seeing (solid lines, red and blue respectively). The MASS/DIMM values are converted to the corresponding observation wavelength with equation \ref{eqn:wave_cal}. The observation wavelength is reported with the date and indicated by each figure's background color: {\em gray} for $R$-band, {\em yellow} for 1000 nm, and no shading for $I$-band. The image FWHM data points each represent the median value of all sources in a single frame. Each panel's caption gives the UT date of observation.} 
  \label{fig:Nightly}
\end{figure*}

\end{document}